# Identification and Separation of Radioactive Isotope Beams by the BigRIPS Separator at the RIKEN RI Beam Factory

N. Fukuda [*], T. Kubo *, T. Ohnishi, N. Inabe, H. Takeda, D. Kameda, H. Suzuki

*RIKEN Nishina Center, RIKEN, 2-1 Hirosawa, Wako, Saitama 351-0198, Japan*

**Abstract**

We have developed a method for achieving excellent resolving power in in-flight particle identification of radioactive isotope (RI) beams at the BigRIPS fragment separator at the RIKEN Nishina Center RI Beam Factory (RIBF). In the BigRIPS separator, RI beams are identified by their atomic number $Z$ and mass-to-charge ratio $A/Q$ which are deduced from the measurements of time of flight (TOF), magnetic rigidity ($B\rho$) and energy loss ($\Delta E$), and delivered as tagged RI beams to a variety of experiments including secondary reaction measurements. High $A/Q$ resolution is an essential requirement for this scheme, because the charge state $Q$ of RI beams has to be identified at RIBF energies such as 200–300 MeV/nucleon. By precisely determining the $B\rho$ and TOF values, we have achieved relative $A/Q$ resolution as good as 0.034% (root-mean-square value). The achieved $A/Q$ resolution is high enough to clearly identify the charge state $Q$ in the $Z$ versus $A/Q$ particle identification plot, where fully-stripped and hydrogen-like peaks are very closely located. The precise $B\rho$ determination is achieved by refined particle trajectory reconstruction, while a slew correction is performed to precisely determine the TOF value. Furthermore background events are thoroughly removed to improve reliability of the particle identification. In the present paper we present the details of the particle identification scheme in the BigRIPS separator. The isotope separation in the BigRIPS separator is also briefly introduced.

*Corresponding authors:

nfukuda@riken.jp (Naoki Fukuda) and kubo@ribf.riken.jp (Toshiyuki Kubo)

# 1. Introduction

In-flight production of radioactive isotope (RI) beams was pioneered in the 1980s at LBNL [1] and GANIL [2] via projectile fragmentation of heavy ion beams. Since the early 1990s, research on exotic nuclei using RI beams was advanced with the construction of in-flight fragment separators worldwide, including RIPS at RIKEN [3], A1200 and A1900 at NSCL/MSU [4, 5], FRS at GSI [6], LISE3 [7] at GANIL. The region of accessible exotic nuclei has been expanded significantly with these facilities. Furthermore the RI beam production via in-flight fission of a uranium beam was pioneered at FRS, demonstrating its excellent features for producing of a wide range of neutron-rich exotic nuclei [8, 9].

In efforts to further expand research on exotic nuclei using RI beams, a new-generation in-flight fragment separator named BigRIPS [10-12] was constructed in March 2007 at the Radioactive Isotope Beam Factory (RIBF) [13] at the RIKEN Nishina Center. The characteristic features of the BigRIPS separator are large ion-optical acceptances and a two-stage structure. The angular acceptances are ±40 mrad horizontally and ±50 mrad vertically, and the momentum acceptance is ±3%, allowing efficient collection of fragments produced by not only projectile fragmentation but also in-flight fission of a $^{238}$U beam. The typical collection efficiency of fission fragments is approximately 50%, even though they are produced with large angular and momentum spreads at RIBF energies. Such large acceptances are realized by the use of superconducting quadrupole magnets with large apertures.

The two-stage structure allows delivery of tagged RI beams and two-stage isotope separation. The first stage of the BigRIPS separator is used for production, collection, and separation of RI beams with an energy degrader, while particle identification of RI beams (separator-spectrometer mode) and/or further isotope separation with another energy degrader (separator-separator mode) are performed in the second stage. The particle identification is based on the TOF-$B\rho$-$\Delta E$ method, in which the time of flight (TOF), magnetic rigidity ($B\rho$), and energy loss ($\Delta E$) are measured to deduce the atomic number ($Z$) and the mass-to-charge ratio ($A/Q$) of RI beams. Such in-flight particle identification is an essential requirement for delivering tagged RI beams, making it possible to perform various types of experiments including secondary reaction measurements. Since the total kinetic energy is not measured in this scheme, and consequently $A$ and $Q$ cannot be determined independently, the resolution in $A/Q$ must be high enough to identify the charge state $Q$ of RI beams. If $Z$ is relatively high, RI beams are not necessarily fully-stripped at RIBF energies such as 200–300 MeV/nucleon. In fact fully-stripped and hydrogen-like peaks often appear in very close positions in a $Z$ versus $A/Q$ particle identification plot. In order to achieve the high $A/Q$ resolution, the ion optics of the second stage is designed with high momentum resolution: The magnification and momentum dispersion at the intermediate focus are 0.927 and 31.7 mm/%, respectively, corresponding to the first-order resolution of 3420 with the object size of 1 mm. Furthermore, the flight path is fairly long

(46.6 m), so that the TOF value can be determined with sufficiently high resolution.

A refined analysis of the measured data plays an important role in precisely determining the $B\rho$ and TOF values, which allows us to realize the high $A/Q$ resolution in our particle identification scheme. The precise $B\rho$ determination is achieved by means of trajectory reconstruction, in which measured particle trajectories are combined with ion-optical transfer matrix elements deduced from experimental data. The TOF value is determined with high resolution by correcting the time with regard to the pulse height (slewing correction). In order to improve reliability of the particle identification, background events are excluded by fully examining detector signals, physical quantities extracted from these signals, and their correlations. Thanks to the achieved excellent particle identification power and low backgrounds, we successfully discovered 45 new isotopes in an experiment conducted in November 2008 [14].

In the present paper, we report on the details of the particle identification scheme in the BigRIPS separator. The isotope separation utilizing the two-stage configuration of the BigRIPS separator is briefly introduced as well.

## 2. Particle identification

*2.1. Outline*

In the TOF-$B\rho$-$\Delta E$ method, we deduce the $Z$ and $A/Q$ values from the measured TOF, $B\rho$, and $\Delta E$ using the equations as follows:

$$\text{TOF} = \frac{L}{\beta c} \tag{1}$$

$$\frac{A}{Q} = \frac{B\rho}{\beta\gamma} \frac{c}{m_u} \tag{2}$$

$$\frac{dE}{dx} = \frac{4\pi e^4 Z^2}{m_e v^2} Nz \left[ \ln\frac{2m_e v^2}{I} - \ln\left(1-\beta^2\right) - \beta^2 \right] \tag{3}$$

Here $L$ is the flight path length, $v$ ($\beta = v/c, \gamma = 1/\sqrt{1-\beta^2}$, where $c$ is the velocity of light) is the velocity of the particle, $m_u = 931.494$ MeV is the atomic mass unit, $m_e$ is the electron mass, and $e$ is the elementary charge. $z$, $N$ and $I$ represent the atomic number, atomic density and mean excitation potential of the material, respectively. $Z$, $A$ and $Q$ represent the atomic, mass and charge (charge state) number of the particle, respectively. The Bethe-Bloch formula [15, 16] shown in Eq. (3), describes the energy loss $\Delta E$.

Fig. 1 shows a schematic layout of the BigRIPS separator along with the standard setup of the beam line detectors used for the particle identification of RI beams. The foci at F3 and F7 are fully achromatic, while those at F1 and F5 are momentum dispersive. As shown in Fig. 1, we measure the TOF by using thin plastic scintillation counters installed at F3 and F7, which are respectively located at the beginning and end of the second stage of the BigRIPS separator. Small beam spots at these achromatic foci allow the TOF measurement with good time resolution. We measure the $\Delta E$ by using a multi-sampling ionization chamber (MUSIC) [17] or a stacked silicon detector installed at F7. The $B\rho$ measurement is made by trajectory reconstruction [18-20] not only in the first half of the second stage (F3–F5) but also the second half (F5–F7), as shown in Fig. 1. For the trajectory reconstruction, we measure particle trajectories, namely positions and angles of fragments, at F3, F5 and F7 by using two sets of position-sensitive parallel plate avalanche counters (PPAC) [21] installed at the respective foci. Ion-optical transfer matrix elements up to the third order, deduced from experimental data, are used for the trajectory reconstruction. The absolute $B\rho$ value of fragments on the central trajectory, called the central $B\rho$ value, is determined by using the magnetic fields of the dipole magnets measured by NMR probes and the central trajectory radii of the dipole magnets deduced from the magnetic field-map data. The TOF and $\Delta E$ measurements are calibrated by using the central $B\rho$ value thus obtained.

As the PPAC detectors and the energy degrader at F5 give rise to energy loss, the twofold $B\rho$ measurement mentioned above is needed to deduce the $A/Q$ value of fragments in combination with the TOF measurement between F3 and F7. In this case, Eqs. (1) and (2) give

$$\text{TOF} = \frac{L_{35}}{\beta_{35}c} + \frac{L_{57}}{\beta_{57}c} \tag{4}$$

$$\left(\frac{A}{Q}\right)_{35} = \frac{B\rho_{35}}{\beta_{35}\gamma_{35}} \frac{c}{m_u} \tag{5}$$

$$\left(\frac{A}{Q}\right)_{57} = \frac{B\rho_{57}}{\beta_{57}\gamma_{57}} \frac{c}{m_u}. \tag{6}$$

Here the subscripts 35 and 57 indicate the quantities related to the F3–F5 and F5–F7 sections, respectively. If the $A/Q$ value does not change at F5, then

$$\frac{\beta_{35}\gamma_{35}}{\beta_{57}\gamma_{57}} = \frac{B\rho_{35}}{B\rho_{57}}. \tag{7}$$

In this case the fragment velocities before ($\beta_{35}$) and after ($\beta_{57}$) F5 can be deduced from Eqs. (4) and (7) using the measured TOF, $B\rho_{35}$ and $B\rho_{57}$ values, allowing the determination of absolute $A/Q$ value. The absolute $Z$ value is derived using the measured $\Delta E$ and $\beta_{57}$ values based on Eq. (3). A two-dimensional plot of $Z$ versus $A/Q$ is used for the particle identification.

We verify the particle identification by detecting delayed $\gamma$-rays emitted from known short-lived isomers by using germanium detectors placed at F7 or other focal planes downstream [14, 22]. The observation of characteristic isomeric γ-rays allows unambiguous isotope identification, a technique known as isomer tagging [23].

We exclude background events, such as those caused by reactions and scatterings, signal pileups and improper detector responses, by checking profiles of beam spot and phase space, consistency of fragment trajectories, various correlation plots made of pulse-height and timing signals in the beam line detectors. This allows us to identify rare events with confidence as well as to improve the reliability of particle identification.

In the following sections, we first describe the details of the background removal, and then those of the particle identification scheme emphasizing the precise $B\rho$ determination by trajectory reconstruction. The procedure for identifying the charge-state changes between F0 and F7 is also described. We show some recent experimental data to illustrate.

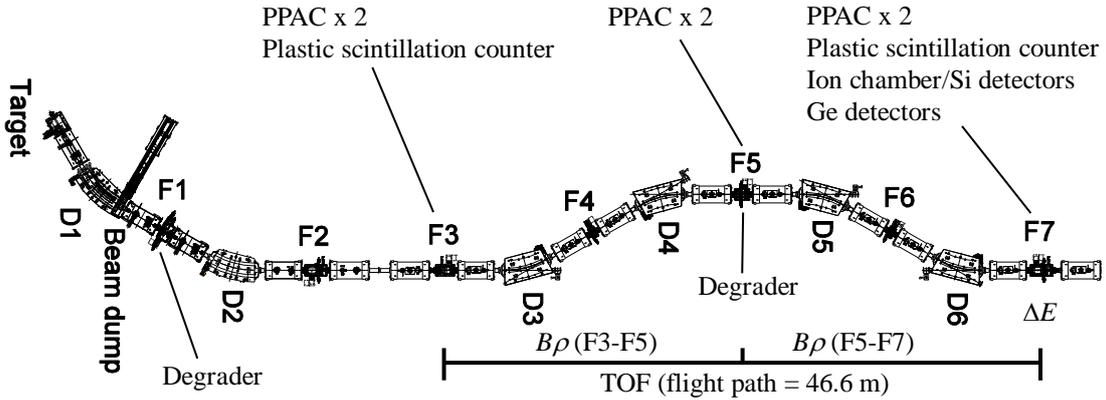

Fig. 1. Schematic layout of the BigRIPS separator. The labels Dn and Fn indicate the positions of dipole magnets and foci, respectively. The first stage includes the components from the production target (F0) to F2, while the second stage spans those from F3 to F7. The positions of the beam-line detectors used for particle identification are also shown along with the measurement scheme of the time of flight (TOF), magnetic rigidity ($B\rho$) and energy loss ($\Delta E$). See text.

*2.2. Removal of background events*

*2.2.1. Background removal by the plastic scintillation counters*

Scintillation signals from the plastic scintillation counter, placed at F3 and F7 for the TOF measurement, are detected by using two photomultiplier tubes (PMT) which are optically coupled to the left and right ends of the detector. The charge-integrated signals from the PMTs contain the

position information of an incident particle, which is expressed as

$$q_1 = q_0 \exp\left(-\frac{L+x}{\lambda}\right) \text{ and } q_2 = q_0 \exp\left(-\frac{L-x}{\lambda}\right), \tag{8}$$

from which we find

$$x = -\frac{\lambda}{2} \ln\left(\frac{q_1}{q_2}\right). \tag{9}$$

Here $q_1$ and $q_2$ represent the signal values obtained from the left and right PMTs, respectively. $\lambda$ denotes the attenuation length of light in the scintillation counter. $L$ and $x$ represent the length of the scintillation counter and the horizontal position of an incident particle, respectively. $q_0$ represents the signal value of the original scintillation. The timing signals from the PMTs also contain the position information of an incident particle which is expressed as

$$x = -\frac{V}{2}(t_2 - t_1). \tag{10}$$

Here $t_1$ and $t_2$ represent the timing information from the left and right PMTs, respectively, while $V$ denotes the propagation speed of light in the scintillation counter. The time difference $t_2 - t_1$ provides the position information.

Eqs. (9) and (10) lead to the following relation:

$$\lambda \ln\left(\frac{q_1}{q_2}\right) = V(t_2 - t_1). \tag{11}$$

A correlation plot between the $\ln(q_1/q_2)$ and $t_2 - t_1$ allows us to remove inconsistent events, because they deviate from the correct correlation given by Eq. (11). Fig. 2 shows an example of the correlation plot.

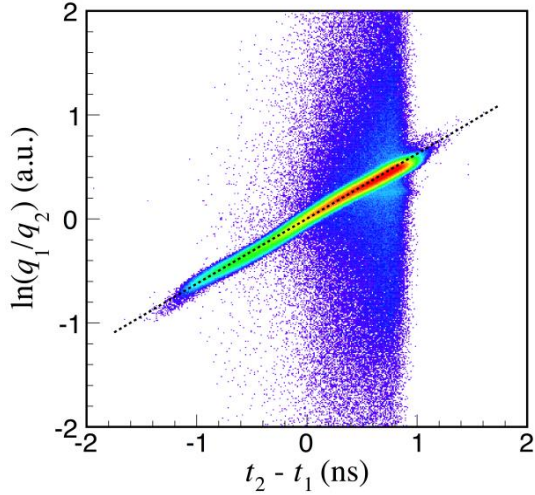

Fig. 2. Example of the background removal by the plastic scintillation counter at F3. A correlation plot between $\ln(q_1/q_2)$ and $t_2 - t_1$ is shown for fission fragments produced by in-flight fission of a $^{238}$U beam at 345 MeV/nucleon. The experimental conditions and the BigRIPS settings are given as G3 setting in Ref. [14]. Normal events follow a linear correlation indicated by the dotted line. See text.

### 2.2.2. Background removal by the PPAC detectors

The position-sensitive PPAC detectors, used for the trajectory measurements at the foci in the BigRIPS separator, adopt a delay-line readout method [21, 24]. The position of an incident particle is determined from the time difference between two timing signals $T_1$ and $T_2$, which are obtained from either ends of the delay line in the PPAC detector. Inconsistent events such as multiple-hit events and $\delta$-ray generation events can be removed by using the sum of these two timing signals expressed as $T_{\text{sum}} = T_1 + T_2$. The $T_{\text{sum}}$ value is a constant independent of the position of an incident particle if the events are normal, and it gets smaller than the normal value if the multiple hits and/or $\delta$-ray generation occur [24].

Furthermore the trajectory measurements using the PPAC detectors allows us to check phase space profiles as well as beam spot profiles of RI beams. Inconsistent events can be excluded by constraining these profiles.

### 2.2.3. Removal of reaction events in the $\Delta E$ measurement

The MUSIC detector used for the $\Delta E$ measurement in the BigRIPS separator consists of twelve anodes and thirteen cathodes aligned alternately. The design is based on Ref. [17]. The electrodes are made of a thin mylar foil aluminized on both sides, and the neighboring anodes are electrically connected in pairs. The six anode signals are read independently and averaged for the $\Delta E$ measurement. The correlation between the neighboring anode signals allows us to identify and remove inconsistent events such as those caused by nuclear reactions in the electrodes and counter gas. The energy loss of fragments varies significantly if the reactions happen in the detector. Similarly, in the case of using a stack of silicon detectors, the correlations between the energy signals from individual detectors are used to remove inconsistent events.

*2.2.4. Background removal by correlation between the detectors at different foci*

The correlations between the detectors installed at different foci also allow us to exclude background events. For instance, inconsistent events generated by nuclear reactions in the plastic scintillation counters at F3 and F7 can be identified by the correlation plot between the charge-integrated signals of these scintillation counters. The correlations between the $\Delta E$ signal from the MUSIC detector and the charge-integrated signals from these scintillation counters also allow the rejection of the nuclear reaction events. Furthermore it is also possible to identify signal pileup events by these correlations, because signal pileups tend to have more effects on the MUSIC detector whose pulse-height signals are slow.

The correlation of measured particle trajectories between two different foci allows us to exclude ion-optically inconsistent events. For instance, we reject the events if the angles measured at F3 and F5 by using the PPAC detectors are not consistent with the ion optics of the BigRIPS separator.

*2.2.5. Removal of events whose charge state changes at F5 focus*

We remove the events whose charge state changes at F5, where the PPAC detectors and the energy degrader are placed, because Eq. (7) is not satisfied and hence the $\beta_{35}$ and $\beta_{57}$ values cannot be derived correctly. Unless such events are removed, the *A/Q* values are not properly calculated, generating background events in the particle identification plot.

In order to identify and remove such charge-state changing events, we also measure the TOF between F3 and F5 ($TOF_{35}$) and between F5 and F7 ($TOF_{57}$) by using the timing information obtained from the PPAC detector at F5. These two TOF measurements allow us to deduce the charge-state numbers before ($Q_{35}$) and after ($Q_{57}$) F5. Fig. 3 shows an example of the charge-state change at F5, where a *Z* versus $Q_{57}/Q_{35}$ plot is shown for fission fragments produced by the in-flight fission of a $^{238}$U beam at 345 MeV/nucleon. The events with $Q_{35} = Q_{57}$ are sufficiently separated from those with $Q_{35} = Q_{57} - 1$, which are located around $Q_{57}/Q_{35} = 1.02$, allowing us to reject the charge-state changing events. Poor time resolution of the PPAC detector is good enough for this purpose: The room-mean-square (rms) resolution is approximately 900 ps. The time resolution of the plastic scintillation counter at F3 and F7 is much better than this value as described in section 2.4.

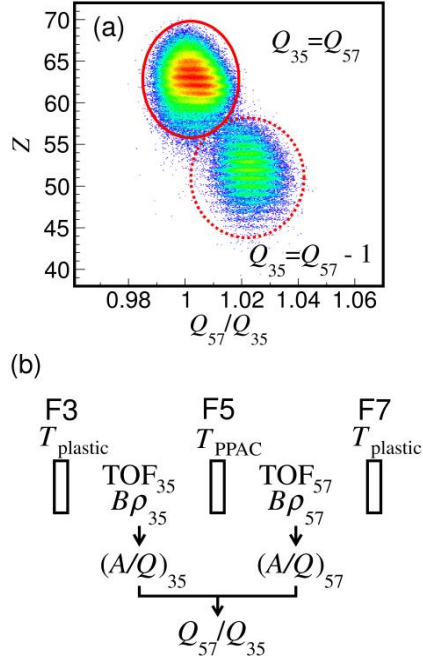

Fig. 3. Example of the charge-state change at F5. (a) $Z$ versus $Q_{57}/Q_{35}$ plot for fission fragments produced in the reaction $^{238}$U + Be (4.9 mm) at 345 MeV/nucleon. The BigRIPS setting is tuned for $^{168}$Gd. The $Q_{35} = Q_{57}$ events are circled with a solid line, while those circled with a dashed line, distributed around $Q_{57}/Q_{35} = 1.02$, correspond to the $Q_{35} = Q_{57} - 1$ events. (b) Diagram conceptually showing the derivation of $Q_{35}$ and $Q_{57}$, where $T_{\text{plastic}}$ and $T_{\text{PPAC}}$ respectively denote the timing information from the plastic scintillation counters and from the PPAC detector. See text.

### 2.2.6. Other methods for background removal

In order to further remove background events, we sometimes utilize the following correlations: $B\rho_{35}$ versus $B\rho_{57}$, TOF$_{35}$ versus TOF$_{57}$, time difference $t_2 - t_1$ of the plastic scintillation counter versus horizontal position measured by the PPAC detector, and $q_1$ versus $q_2$ and $t_1$ versus $t_2$ in the plastic scintillation counters.

Signal pileup events are also electrically identified.

### 2.3. Particle trajectory reconstruction

Here we describe the details of our trajectory reconstruction scheme, which allows the precise $B\rho$ determination and hence the high $A/Q$ resolution in the particle identification. Fig. 4 shows an example of the $Z$ versus $A/Q$ particle identification plot for fission fragments produced using in-flight fission of a 345 MeV/nucleon $^{238}$U beam. In this measurement we search for new isotopes in the neutron-rich region [14]. The experimental conditions are given as G3 setting in Ref. [14].

Thanks to the high *A/Q* resolution that we achieved by trajectory reconstruction, the fully-stripped and hydrogen-like peaks are clearly separated in the particle identification plot, as seen in Fig. 4.

*2.3.1. Ion-optical transformation and trajectory reconstruction*

The trajectory reconstruction is performed by combining the measured fragment trajectories with the ion-optical transformation between the initial and the final coordinate vectors [25, 26]. In the case of the horizontal plane between F3 and F5, the ion-optical transformation up to the second order is given by

$$\begin{cases} x_5 = (x|x)x_3 + (x|a)a_3 + (x|\delta)\delta_{35} \\ \quad + (x|xx)x_3^2 + (x|xa)x_3 a_3 + (x|x\delta)x_3 \delta_{35} + (x|aa)a_3^2 + (x|a\delta)a_3 \delta_{35} \\ \quad + (x|\delta\delta)\delta_{35}^2 + (x|yy)y_3^2 + (x|yb)y_3 b_3 + (x|bb)b_3^2 \\ a_5 = (a|x)x_3 + (a|a)a_3 + (a|\delta)\delta_{35} \\ \quad + (a|xx)x_3^2 + (a|xa)x_3 a_3 + (a|x\delta)x_3 \delta_{35} + (a|aa)a_3^2 + (a|a\delta)a_3 \delta_{35} \\ \quad + (a|\delta\delta)\delta_{35}^2 + (a|yy)y_3^2 + (a|yb)y_3 b_3 + (a|bb)b_3^2 \end{cases} \quad (12)$$

Here the notation is based on the COSY INFINITY code [27]. The coordinates ($x_3$, $a_3$) and ($x_5$, $a_5$) represent the positions and angles in the horizontal plane at F3 and F5, respectively. The $\delta_{35}$ denotes the fractional $B\rho$ deviation from the central value $B\rho_0$, which is expressed as $(B\rho - B\rho_0)/B\rho_0$. The coefficients $(x|x)$, $(x|a)$, $(x|\delta)$, … $(a|bb)$ in the equations are called transfer matrix elements or transfer maps. The first-order matrix elements $(x|x)$, $(x|a)$ and $(x|\delta)$ represent the image magnification, the angular dependence and the momentum dispersion, respectively. The angular dependence $(x|a)$ equals zero if the focusing is realized.

In our trajectory reconstruction, the quantities $\delta_{35}$ and $a_3$ are determined (reconstructed) using Eq. (12) from the measured positions and angles at F3 and F5: ($x_3$, $a_3$) and ($x_5$, $a_5$). Furthermore, in order to improve the $B\rho$ resolution as much as possible, we use the transfer matrix elements determined from experimental data. We solve Eq. (12) numerically by means of the Newton-Raphson method and obtain the reconstructed $\delta_{35}$ and $a_3$. The obtained $\delta_{35}$ allows us to calculate the corresponding $B\rho_{35}$ value using the measured central $B\rho$. The comparison between the measured and reconstructed $a_3$ values allows us to verify the correctness of our trajectory reconstruction. The ion-optical transformations between F5 and F7 are given similarly to Eq. (12) and the trajectory reconstruction is performed in a similar way.

*2.3.2. Determination of first-order transfer matrix elements*

Our trajectory reconstruction procedure begins with the determination of the first-order transfer matrix elements, followed by an improvement process in which higher-order transfer matrix elements are introduced. We experimentally derive the first-order matrix elements from the measured various correlations between the initial and final foci, such as position versus position, position versus angle, angle versus angle, position versus TOF, and angle versus TOF correlations. For example, the gradient of the correlation between the positions $x_5$ and $x_3$ gives the $(x|x)$ element if the events with $a_3 \approx 0$ and $\delta_{35} \approx 0$ are selected, because Eq. (12) can be reduced to $x_5 \approx (x|x)x_3$ under these conditions. Here we select specific isotopes whose events provide the most suitable conditions to deduce the matrix element. Such event selections are performed based on a *Z* versus *A/Q* plot and $B\rho$ values which are preliminarily obtained using the calculated first-order matrix elements with the code COSY INFINITY. Other elements such as $(x|a)$, $(a|x)$, and $(a|a)$ can be obtained in a similar way. The correlation between $x_5$ and $\delta_{35}$ allows us to determine the $(x|\delta)$ element, because $x_5 \approx (x|\delta)\delta_{35}$ if the events with $x_3 \approx 0$ and $a_3 \approx 0$ are selected. Here we use the $\delta_{35}$ values obtained from the TOF measurement.

Fig. 5 shows an example of the experimental derivation of the first-order matrix elements between F3 and F5. The obtained matrix elements are summarized in Table 1 along with those calculated [28] with the COSY INFINITY code. The comparison reveals that some matrix elements, such as $(x|a)$, deviate from the calculations to a significant extent. If the focusing condition is realized well, this matrix element should be zero like the calculated value. The finite non-zero value of $(x|a)$, listed in Table 1, corresponds to a 0.12% shift in $\delta_{35}$ for $a_3 = 20$ mrad (a typical angular spread of fission fragments in the second stage of the BigRIPS separator). Such an effect results in a non-negligible deterioration of the $B\rho$ resolution, indicating the importance of using the transfer matrix elements obtained from experimental data.

In case of the first-order transfer matrix between F5 and F7, we first determine the inverse matrix elements from F7 to F5 experimentally, and then invert them to derive the forward transfer matrix elements. In case of the inverse transfer matrix, the above-mentioned event selections can be made independently of the $\delta$ value, because the F7 focus is fully achromatic. Otherwise it is difficult to derive the $(x|\delta)$ and $(a|\delta)$ elements.

*2.3.3. Determination of higher-order transfer matrix elements*

Higher-order transfer matrix elements are determined by the empirical method, in which their values are empirically adjusted so as to improve the *A/Q* resolution. As long as the trajectory reconstruction using Eq. (12) is properly made, the deduced *A/Q* value for any isotope should be independent of the position, angle, and fractional $B\rho$ deviation. Such conditions allow us to achieve the ion-optically best possible *A/Q* resolution. Fig. 6 shows examples of the *A/Q* versus $x_3$, $a_3$, and $\delta_{35}$ plots for Sn isotopes produced by the in-flight fission. The plots shown in Fig. 6 (a), (c), and (e) are those using only the first-order matrix elements determined from experimental data, in which some dependence on the $x_3$, $a_3$, and $\delta_{35}$ are clearly seen. Fig. 6 (b), (d), and (f) show the plots when the second- and third-order matrix elements are introduced and adjusted empirically, so that the dependence can be canceled out. In these plots the *A/Q* value is much less dependent on $x_3$, $a_3$, and $\delta_{35}$, indicating that higher-order elements are properly derived and the *A/Q* resolution is improved significantly. Fig. 7 shows the comparison between the measured and reconstructed $a_3$ angles at F3. Their small difference confirms the correctness of our present method. The higher-order matrix elements thus obtained are summarized in Table 2. Note that we assume the dependence seen in Fig. 6 (a), (c), and (e) is merely caused by the fact that the possible higher-order terms are not included.

We have derived some higher-order matrix elements by using the same method as in the derivation of the first-order matrix elements. In this case, however, the trajectory reconstruction is not so successful, probably because the obtained higher-order matrix elements are not accurate enough.

### *2.3.4. Improvement of A/Q resolution by trajectory reconstruction*

The *Z* versus *A/Q* particle identification plot shown in Fig. 4 has been obtained using the experimentally derived matrix elements listed in Table 1 and 2. Fig. 8 (a), (b), and (c) show the *A/Q* spectra of Sn isotopes produced in the same run as in Fig. 4. These *A/Q* spectra have been obtained using the following three different transfer matrix elements in the trajectory reconstruction: (a) the calculated first-order matrix elements (see Table 1), (b) the experimentally derived first-order matrix elements (see Table 1), and (c) the experimentally derived matrix elements up to the third order (see Table 1 and 2). The relative rms *A/Q* resolution in Fig. 8 (c) is as high as σ = 0.038%, revealing the significant improvement compared to the other two spectra and hence the importance of the trajectory reconstruction including the high-order matrix elements. Furthermore the degree of tail separation between different charge states also reveals significant improvement, indicating the advantages of including the higher-order matrix elements. Such improvement is crucial to reliably identify isotopes from a small number of events, which often happens in experiments to search for new isotopes.

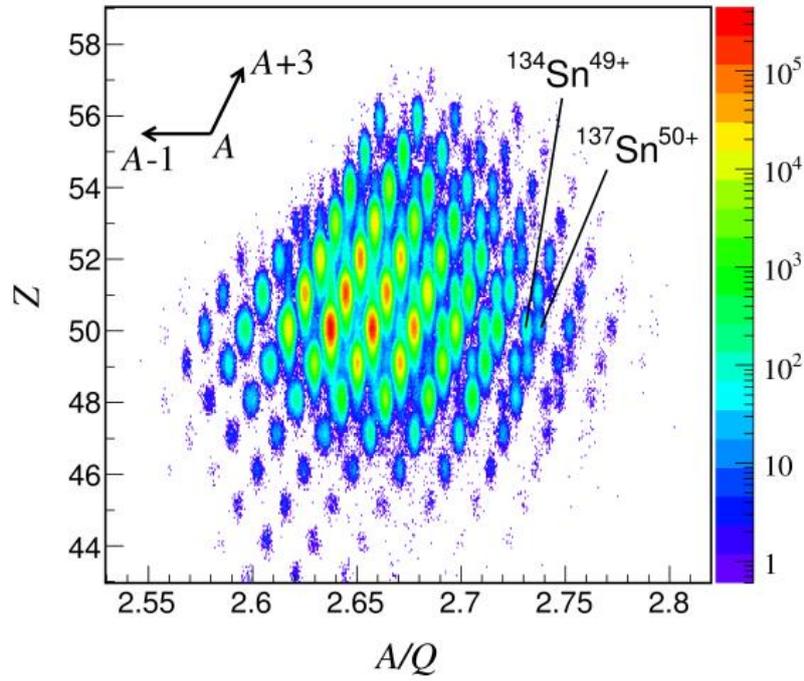

Fig. 4. *Z* versus *A/Q* particle identification plot for fission fragments produced in the $^{238}$U + Pb reaction at 345 MeV/nucleon. The experimental conditions and the BigRIPS settings are given as G3 setting in Ref. [14]. The fully stripped ion $^{137}$Sn$^{50+}$ and the hydrogen-like ion $^{134}$Sn$^{49+}$ are labeled to demonstrate the high *A/Q* resolution achieved by the trajectory reconstruction. See Text.

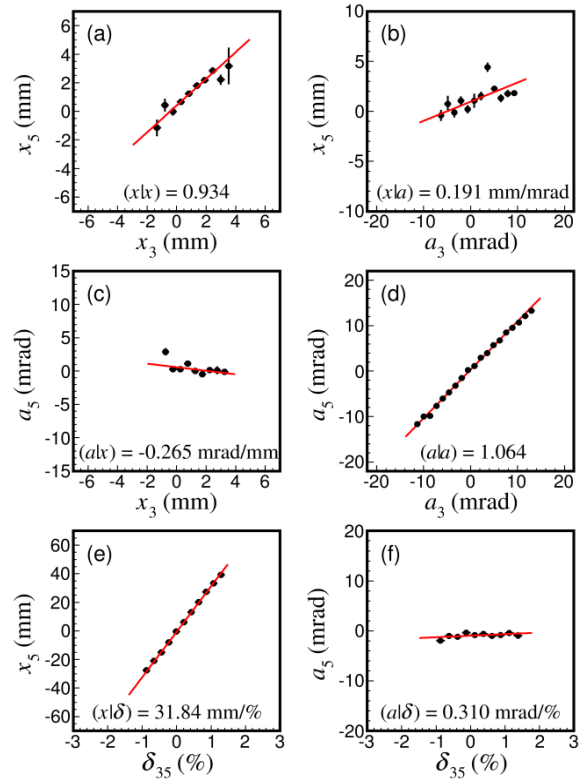

Fig. 5. Example of the first-order matrix element derivation. The derivation is made with the selected isotope $^{141}$I$^{53+}$ which is produced by in-flight fission of a $^{238}$U beam at 345 MeV/nucleon. The settings of the BigRIPS second stage are the same as those of Fig. 4, while the settings of the first stage are adjusted so that the experimental conditions can be optimized for the derivation of the matrix elements. (a) Plot of the correlation between $x_5$ and $x_3$ for $a_3 \approx 0$ and $\delta_{35} \approx 0$. The red solid line shows the fitted line using a linear function. The matrix element $(x|x)$ can be obtained from the gradient of the correlation using $x_5 \approx (x|x)x_3$. Other elements, such as $(x|a)$, $(a|x)$, and $(a|a)$, can also be derived from the correlations between (b) $x_5$ and $a_3$, (c) $a_5$ and $x_3$, and (d) $a_5$ and $a_3$ in a similar way. (e) Plot of the correlation between of $x_5$ and $\delta_{35}$ for $x_3 \approx 0$ and $a_3 \approx 0$. The $\delta_{35}$ value is calculated from the TOF value. The $(x|\delta)$ element can be derived from $x_5 \approx (x|\delta)\delta_{35}$, and $(a|\delta)$ can be derived from (f) the correlation between $a_5$ and $\delta_{35}$ in a similar way. See text.

Table 1 First-order matrix elements derived from experimental data. The same matrix elements calculated using the code COSY INFINITY are shown for comparison.

| Matrix elements | Experimentally derived | Calculated using COSY INFINITY |
|:---:|:---:|:---:|
| $(x\|x)$ | $0.934 \pm 0.094$ | 0.927 |
| $(a\|x)$ | $-0.265 \pm 0.138$ | -0.020 |
| $(x\|a)$ | $0.191 \pm 0.039$ | -0.005 |
| $(a\|a)$ | $1.064 \pm 0.009$ | 1.079 |
| $(x\|\delta)$ | $31.84 \pm 0.090$ | 31.67 |
| $(a\|\delta)$ | $0.310 \pm 0.209$ | 0.015 |
| Determinant | $1.044 \pm 0.01$ | 1 |

The units of $x$, $a$, and $\delta$ are given in mm, mrad, and %, respectively.

Table 2 Second- and third-order transfer matrix elements empirically determined from experimental data.

| Matrix elements | Empirically derived |
|:---:|:---:|
| $(x\|xa)$ | 0.008 |
| $(x\|x\delta)$ | -0.065 |
| $(x\|a\delta)$ | -0.051 |
| $(x\|bb)$ | 0.003 |
| $(x\|aaa)$ | 0.00009 |
| $(a\|xa)$ | 0.0027 |
| $(a\|x\delta)$ | -0.011 |
| $(a\|aa)$ | -0.0058 |
| $(a\|a\delta)$ | 0.0065 |
| $(a\|aa\delta)$ | -0.0018 |

The units of $x$, $a$, $b$, and $\delta$ are given in mm, mrad, mrad, and %, respectively.

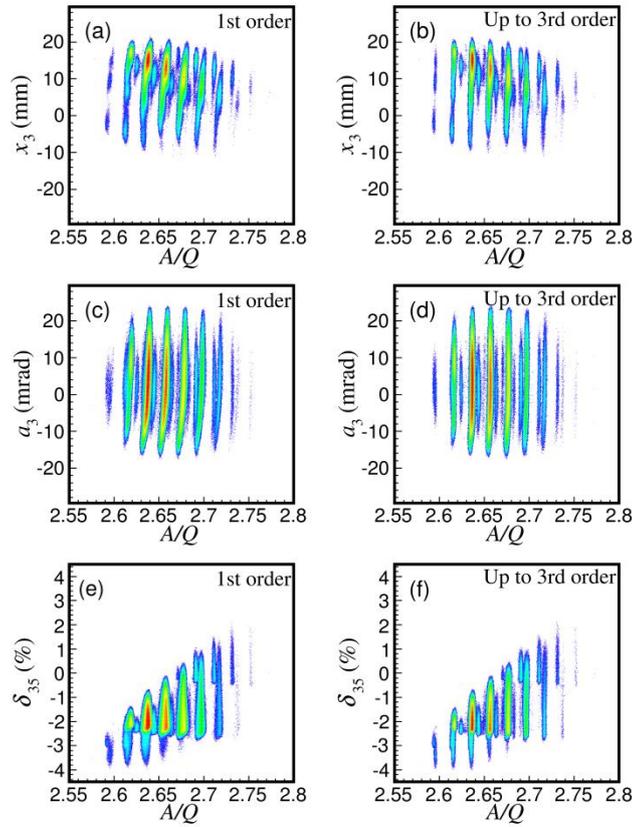

Fig. 6. Example of the higher-order matrix element derivation. The deviation is shown for the Sn isotopes which are produced by in-flight fission of a $^{238}$U beam at 345 MeV/nucleon. The experimental conditions and the BigRIPS settings are the same as those of Fig. 4. Left panels: The correlations between (a) $x_3$ and $A/Q$, (c) $a_3$ and $A/Q$, and (e) $\delta_{35}$ and $A/Q$ when the trajectory reconstruction is performed using the first-order matrix elements derived from experimental data. The $A/Q$ values depend on $x_3$, $a_3$, and $\delta_{35}$ because higher-order matrix elements are not included. Right panels: The correlations between (b) $x_3$ and $A/Q$, (d) $a_3$ and $A/Q$, and (e) $\delta_{35}$ and $A/Q$ when the trajectory reconstruction is performed by including the experimentally derived higher-order matrix elements up to the third order. These plots reveal that the residual dependence is almost completely removed and that the higher-order matrix elements are derived correctly by the empirical method. See text.

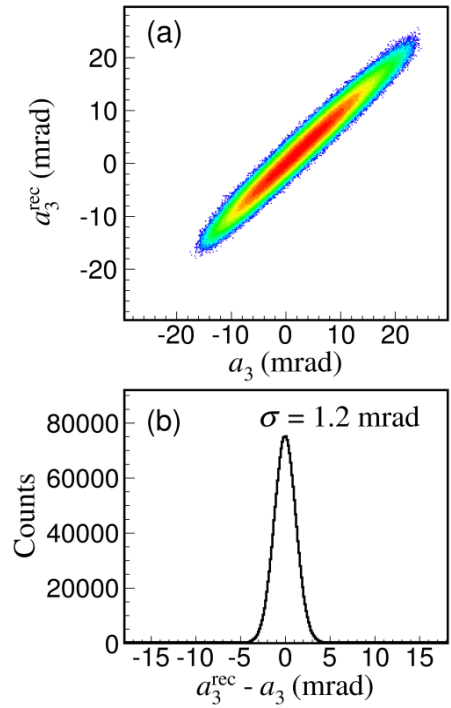

Fig. 7. Example of the verification of trajectory reconstruction. The verification is shown for fission fragments produced by in-flight fission of a $^{238}$U beam at 345 MeV/nucleon. The experimental conditions and the BigRIPS settings are the same as those of Fig. 4. (a) Two-dimensional plot of the reconstructed and measured angles ($a_3$) at F3. (b) Distribution of the difference between the reconstructed and measured angles. See text.

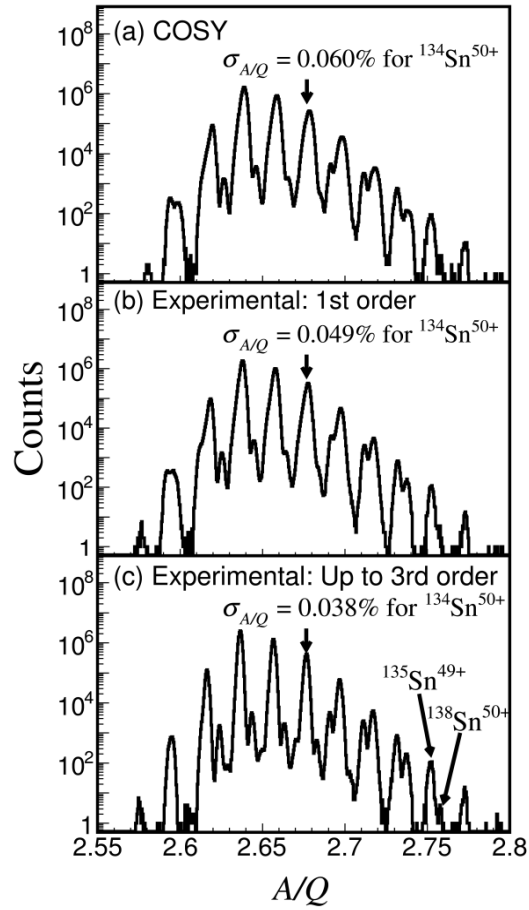

Fig. 8. Comparison of the *A/Q* resolution among three different transfer matrix elements used in the trajectory reconstruction. The comparison is shown for Sn isotopes produced by in-flight fission of a $^{238}$U beam at 345 MeV/nucleon. The experimental conditions and the BigRIPS settings are the same as those of Fig. 4. (a) *A/Q* spectrum obtained using the first-order matrix elements calculated by the COSY INFINITY code, (b) using the experimentally derived first-order matrix elements, and (c) using the experimentally derived matrix elements up to the third order. The relative rms *A/Q* resolution for $^{134}$Sn$^{50+}$ is shown as $\sigma_{A/Q}$ in the figure. See text.

### 2.4. Determination of TOF

The absolute calibration of the TOF measurement is made using the measured central $B\rho$ value of the BigRIPS separator, which allows us to calculate the TOF value between F3 and F7. We use RI beams as well as primary beams for the TOF calibration. When using RI beams, preliminary particle identification is needed to calculate the TOF value. The isomer tagging method is often used for this purpose.

Leading-edge discriminators are used for the time measurement by the plastic scintillation counters at F3 and F7. In order to improve the TOF resolution and hence the *A/Q* resolution, we carry out a slewing correction for the time measurement, in which the jitter of the time signal is corrected using the charge-integrated signal from the scintillation counter. In our slewing correction method, each time signal is corrected using the following equations [29, 30]:

$$\tau = t + \Delta t_{\text{slew}}$$
$$\Delta t_{\text{slew}} = \frac{c_{\text{slew}}}{\sqrt[p]{q}}. \quad (13)$$

Here $\tau$ is the actual arrival time, $t$ is the observed arrival time, and $q$ is the integrated charge signal. The slewing effect term $\Delta t_{\text{slew}}$ depends on the two parameters $c_{\text{slew}}$ and $p$. In our analysis, we empirically determine these two parameters such that the highest time resolution is achieved.

The typical rms time resolution is approximately 40 ps, which corresponds to a relative TOF resolution of 0.017% for a 300 MeV/nucleon particle ($\beta$ = 0.65). Fig. 9 demonstrates the improvement of the *A/Q* resolution achieved by the TOF slewing correction.

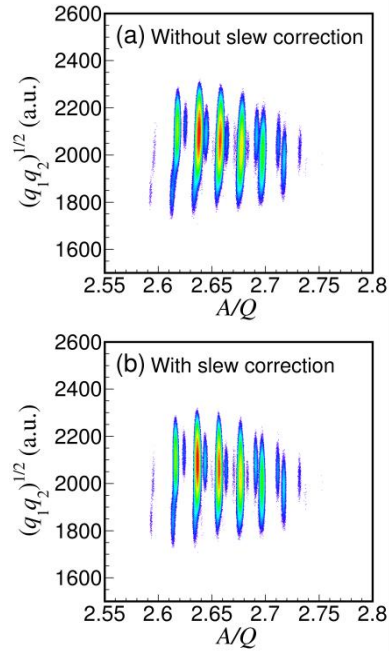

Fig. 9. Improvement of the *A/Q* resolution by the TOF slewing correction which is shown for Sn isotopes produced by in-flight fission of a $^{238}$U beam at 345 MeV/nucleon. The experimental conditions and the BigRIPS settings are the same as those of Fig. 4. (a) and (b) show a two-dimensional plot of $(q_1 q_2)^{1/2}$ versus *A/Q* without and with the slewing correction, respectively. The $(q_1 q_2)^{1/2}$ represents a geometrical mean of the signals $q_1$ and $q_2$, which are obtained from the left and right ends of the plastic scintillation counter, respectively. The geometrical mean cancels out the position dependence of the signals. See text.

*2.5. Determination of the atomic number*

The atomic number Z is determined based on Eq. (3) from the Δ*E* value measured at F7 by the MUSIC detector and the $\beta_{57}$ (= $v_{57}/c$) value determined by the TOF and twofold $B\rho$ measurements. As described in Section 2.2.3, a total of six anode signals are obtained from the MUSIC detector, and the geometrical average of these signals is used as the Δ*E* value to achieve the best possible resolution. The absolute calibration of the Δ*E* measurement is made based on the energy loss calculation by ATIMA [31], for which the measured central $B\rho$ value of the BigRIPS separator is used similarly to the TOF measurement.

In practice, we calculate the absolute Z value by using the following equation [32] containing the two calibration parameters $C_1$ and $C_2$:

$$Z = C_1 v_{57} \sqrt{\frac{\Delta E}{\ln \frac{2 m_e v^2}{I} - \ln\left(1 - \beta_{57}^2\right) - \beta_{57}^2}} + C_2. \quad (14)$$

Here we empirically determine the $C_1$ and $C_2$ parameters according to the Z identification made by the isomer tagging method.

*2.6. Best particle identification power achieved so far*

We optimize the particle identification by iteratively improving the *A/Q* resolution and reducing the number of background events as described in the preceding sections. The derivation of transfer matrix elements and the TOF slew correction are iteratively performed such that the *A/Q* resolution is best optimized. Fig. 10 shows an example of the best-optimized particle identification plot, which was used to identify neutron-rich new isotopes produced using in-flight fission of a 345 MeV/nucleon $^{238}$U beam [14]: (a) the Z versus *A/Q* plot and (b) the *A/Q* spectrum of Rh isotopes (Z = 45). Here the achieved absolute rms *A/Q* resolution is as high as $9.2 \times 10^{-4}$, corresponding to a relative rms *A/Q* resolution of 0.034%. In case of the Rh isotopes we could identify the new isotopes $^{123,124,125,126}$Rh, as labeled in the figure.

As seen in Fig. 10, the absolute rms *A/Q* resolution ($\sigma_{A/Q}$) is much better than the peak separation between the fully-stripped and hydrogen-like peaks which are located very closely in the *A/Q* spectrum, allowing the clear identification of the new isotopes including those with low statistics. For instance, the *A/Q* difference between $^{123}$Rh$^{45+}$ and $^{120}$Rh$^{44+}$ (neighboring hydrogen-like ions) is $5.6 \times 10^{-3}$ corresponding to $6.1 \sigma_{A/Q}$. The good tail separation and the low backgrounds also help the clear identification. Furthermore the centroids of the observed *A/Q* peaks agree well with the calculation using mass values, indicating the accuracy of the calibration. The deviation is less than $1.0 \times 10^{-3}$ in terms of the absolute *A/Q* value, which is comparable to the $\sigma_{A/Q}$ value. This also helps us to achieve the reliable identification of the new isotopes, especially when the statistics are low.

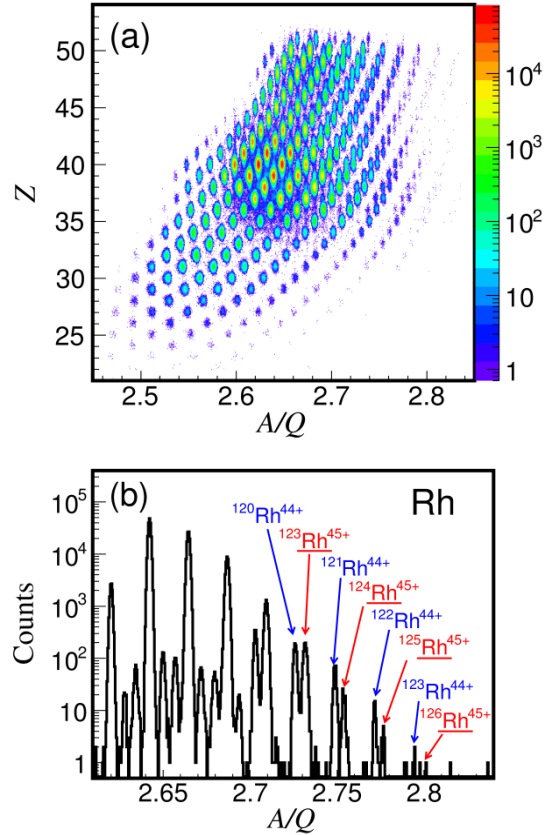

Fig. 10. (a) $Z$ versus $A/Q$ particle identification plot for fission fragments produced in the $^{238}$U + Be reaction at 345 MeV/nucleon. The experimental conditions and the BigRIPS settings are given as G2 setting in Ref. [14]. (b) $A/Q$ spectrum of Rh isotopes ($Z = 45$) produced in the same reaction. Some fully-stripped and hydrogen-like peaks are labeled in the $A/Q$ spectrum. The new isotopes which we identified are indicated by the underline. See text.

## 2.7. *Changes of charge state*

The charge state of RI beams sometimes changes while traveling through the BigRIPS separator, because they pass through the materials placed at the foci, such as the energy degraders and the detectors. Such charge-state changes can be investigated by the correlation plot between the horizontal position $x_3$ at the F3 focus and the $A/Q$ value of RI beams determined in the second stage of the BigRIPS separator.

Fig. 11 (b) shows an example of the $x_3$ versus $A/Q$ correlation plot for Sn isotopes produced by in-flight fission of a $^{238}$U beam, where the charge-stage changes are labeled using the three different lines and the solid circles and triangles. The diagram in Fig. 11 (c) illustrates some details of the labeling. Fig. 11 (a) shows the corresponding *A/Q* spectrum. The closed circles and triangles indicate the isotopes whose charge states are identified as 50+ (fully-stripped) and 49+ (hydrogen-like) in the second stage, respectively. As demonstrated in the preceding sections, the excellent *A/Q* resolution allows us to identify the charge state *Q* in the second stage. A group of relatively prominent peaks labeled by the circles and the solid line correspond to the isotopes which travel throughout the separator as fully-stripped 50+ ions. Note that the most probable charge state is 50+ for Sn isotopes at RIBF energies. Those labeled by the triangles and the solid line correspond to the isotopes which are fully stripped between the production target (F0) and F3 and then change their charge state to 49+ by picking up an electron. In these two cases, isotopes with the same mass number are located at the same $x_3$ position, although they have different *A/Q* values in the second stage.

The displacement in $x_3$ depends on the energy loss by the energy degrader at F1 and the mass dispersion at F3. A group of the peaks labeled by the dashed and dotted lines in Fig. 11 (b) corresponds to the isotopes whose charge states are 49+ and 48+ between F0 and F3, respectively. Because they have the same *Bρ* value after the analysis by the first dipole magnet of the BigRIPS separator, these 49+ and 48+ isotopes are slower than those fully-stripped, resulting in larger energy loss in the energy degrader at F1. This causes the position displacement at F3 for the 49+ and 48+ isotopes as shown in Fig. 11 (b), allowing us to identify the charge states. As labeled by the closed circles and also shown in Fig. 11 (c), these 49+ and 48+ isotopes lose one and two electrons at F3, respectively, and identified as 50+ ions in the second stage. There are also some isotopes whose charge state changes at the F1 degrader. However they are removed by the slits at the F2 focus, because the resulting *Bρ* change is approximately 2% or larger. This way we know how the charge state changes in the BigRIPS separator, which is helpful to check simulation calculations for transmission of RI beams.

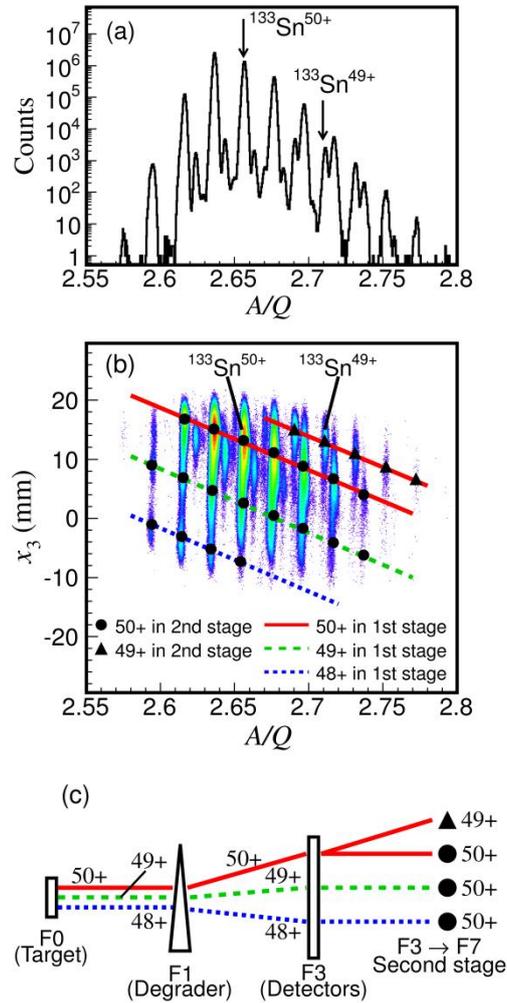

Fig. 11. Example of the charge-state changes which are shown for Sn isotopes produced by in-flight fission of a 345 MeV/nucleon $^{238}$U beam. The experimental conditions and the BigRIPS settings are the same as those of Fig. 4. (a) $A/Q$ spectrum. (b) Correlation plot between the horizontal position at F3 ($x_3$) and the $A/Q$ value determined in the second stage. The closed circles and triangles indicate the isotopes whose charge states are identified as 50+ (fully-stripped) and 49+ (hydrogen-like) in the second stage of the BigRIPS separator, respectively. The closed circles are further classified into different groups by the solid, dashed, and dotted lines, which indicate the isotopes that travel from F0 to F3 as 50+, 49+, and 48+ ions, respectively, without changing their charge states at F1. The group labeled by the triangles and the solid line corresponds to the isotopes that are fully stripped between F0 and F3 and then change their charge state to 49+ by picking up an electron at F3. The $^{133}$Sn$^{50+}$ and $^{133}$Sn$^{49+}$ ions are labeled as an example. They are located at the same position at F3. (c) Diagram showing these charge-state changes. See text.

## 3. Two-stage isotope separation

The second stage in the BigRIPS separator, while acting as a spectrometer for the particle identification, is often used to further separate the RI beam by placing the second energy degrader at the F5 focus. Since isotope separation using an energy degrader depends on the energy of RI beams, we choose the thickness of the energy degraders at the F1 and F5 foci so as to optimize the purity of RI beams. This two-stage separation [10, 33], which combines the isotope separation in the first and second stages, is also effective when unwanted isotopes are transmitted as contaminants due to the charge state change or the secondary reactions in the first degrader at F1. Here we introduce two typical examples.

Fig. 12 shows a $Z$ versus $A/Q$ plot of fission fragments for the following three cases: (a) no energy degraders at F1 and F5, (b) an energy degrader only at F1, and (c) energy degraders at both F1 and F5. Here the fission fragments were produced in the reaction $^{238}$U + Pb at 345 MeV/nucleon. As seen in Fig. 12 (b), the charge-state changes at the F1 degrader produces a formidable amount of contaminants in the region of $Z = 30$–$45$. These contaminants are efficiently removed by the second degrader at F5, as shown in Fig. 12 (c).

Another example shows the case in which reaction products generated at the F1 degrader are removed by the second degrader at F5. Fig. 13 shows a $\Delta E$ versus TOF particle identification plot for fragments produced by projectile fragmentation of a 345 MeV/nucleon $^{48}$Ca beam. The three cases are shown in the same way as in Fig. 12. The reaction products are efficiently removed by the second degrader, as shown in Fig. 13 (b) and (c).

We emphasize that our particle identification method can be used without any significant deterioration in the particle identification power, even if we use a thick energy degrader at F5, which results in large energy loss.

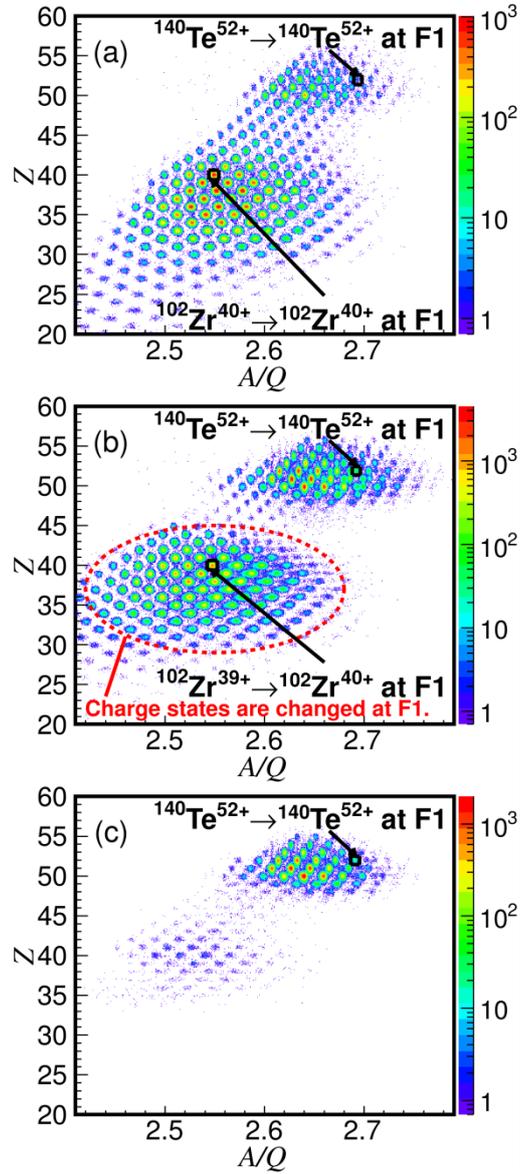

Fig. 12. $Z$ versus $A/Q$ plot for fission fragments produced in the reaction $^{238}$U beam + Pb (1.5 mm) at 345 MeV/nucleon. The $B\rho$ setting of the BigRIPS separator is tuned for $^{140}$Te$^{52+}$. (a) No energy degraders are used at F1 and F5. (b) An energy degrader (Al 3 mm) is used only at F1. (c) Energy degraders are used at both F1 (Al 3 mm) and F5 (Al 1.8 mm). As an example, $^{142}$Te and $^{102}$Zr are labeled by open rectangles along with the charge states before and after the F1 degrader. The charge-state changes at the F1 degrader produce a significant amount of contaminants in the region of $Z$ = 30–45, as circled by the dashed line in (b). See text.

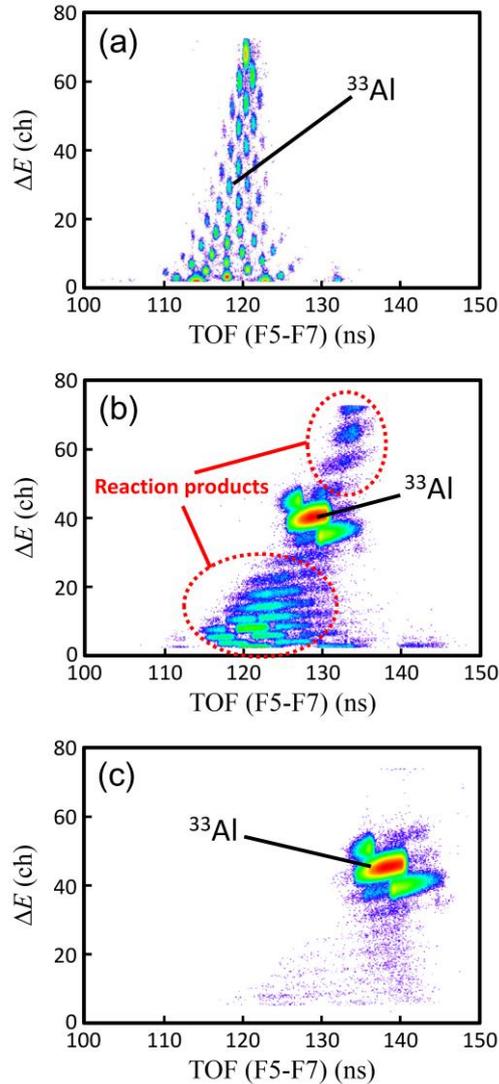

Fig. 13. $\Delta E$ versus TOF particle identification plot for fragments produced by the projectile fragmentation of a 345 MeV/nucleon $^{48}$Ca beam on a 10-mm thick Be target. The BigRIPS settings are tuned for the production of $^{33}$Al.   (a) No energy degraders are used at F1 and F5.   (b) An energy degrader (Al 15 mm) is used only at F1. The secondary reactions in the first energy degrader at F1 generate a significant amount of background events as circled by the dashed lines. (c) Energy degraders are used at both F1 (Al 15 mm) and F5 (Al 10 mm). See text.

## 4.  Conclusion

We have developed a method to achieve excellent *A/Q* resolution in the in-flight particle identification of RI beams, which is routinely used at the BigRIPS fragment separator at RIKEN RIBF. The achieved resolution is high enough to clearly identify the charge state *Q* in the *Z* versus *A/Q* particle identification plot, where fully-stripped and hydrogen-like peaks appear in very closely located positions. Such performance is essential at the RIBF energies where fragments are not necessarily fully-stripped.

We have achieved the high *A/Q* resolution by precisely determining the $B\rho$ and TOF values. The precise $B\rho$ determination has been realized by the trajectory reconstruction method for which ion-optical transfer matrix elements are experimentally determined up to the third-order. The precise TOF determination has been realized by the slew correction method for the TOF signals. We iteratively perform the derivation of transfer matrix elements and the slew correction such that the *A/Q* resolution is best optimized. Furthermore we thoroughly remove background events to enhance the reliability of particle identification.

The excellent particle identification thus achieved has allowed us to deliver tagged RI beams to a variety of experiments at RIBF. Furthermore it helps us to reliably identify new isotopes from a very small number of events. Such an enhanced capability of the BigRIPS separator is significantly advancing the research on exotic nuclei at RIBF. Other new-generation in-flight RI-beam facilities, including the Super-FRS at GSI FAIR [33] and the FRIB at MSU [34], are presently being developed with the same goal.

**Acknowledgements**

The measurements in the present paper were carried out at the RI Beam Factory operated by RIKEN Nishina Center, RIKEN and CNS, University of Tokyo. The authors are grateful to the RIBF accelerator crew for providing the primary beams. The authors would like to thank Dr. Y. Yano, RIKEN Nishina Center, for his support and encouragement. T. K. is grateful to Dr. J. Stasko for his careful reading of the manuscript.